\documentstyle[twocolumn,epsfig,prl,aps]{revtex}
\begin{document}
\draft
\newcommand{\kpkpee}{\emph{$K_{\pi e e}$}}
\newcommand{\kppp}{\emph{$K_{\tau}$}}
\newcommand{\kpp}{$K^{+}\rightarrow\pi^{+}\pi^{0}$}
\newcommand{\peeg}{$\pi^{0}\rightarrow e^{+}e^{-}\gamma$ }
\newcommand{\penu}{$\pi^{-}\rightarrow e^{-}\overline{\nu}$}
\newcommand{\pmunu}{$\pi^{+}\rightarrow \mu^{+}\nu$}
\newcommand{\kdal}{\emph{$K_{Dal}$}}
\newcommand{\kpee}{$K^{+}\rightarrow\pi^{+}e^+e^-$}
\newcommand{\kpme}{$K^{+}\rightarrow\pi^{+}\mu^+e^-$}
\newcommand{\kpimumu}{$K^{+}\rightarrow\pi^{+}\mu^+\mu^-$}
\newcommand{\kpill}{$K^{+}\rightarrow\pi^{+}l^+l^-$}
\newcommand{\ktau}{$K^{+}\rightarrow\pi^{+}\pi^+\pi^-$}
\newcommand{\kpimue}{$K^{+}\rightarrow\pi^{+}\mu^+e^-$}
\newcommand{\dal}{$K_{dal}$}
\newcommand{\pee}{$K_{\pi e e}$}
\newcommand{\pmm}{$K_{\pi \mu\mu}$}
\newcommand{\taus}{$K_{\tau}$} 
\newcommand{\pfour}{${\cal O}(p^4)$}
\newcommand{\psix}{${\cal O}(p^6)$}
\wideabs{
\title{
A New Measurement of the Rare Recay $K^+\rightarrow \pi^+ \mu^+ \mu^-$
}
\author{
H.~Ma$^1$, 
R.~Appel$^{6,3}$, G.~S.~Atoyan$^4$, B.~Bassalleck$^2$,  
D.~R.~Bergman$^6$\cite{DB}, N.~Cheung$^3$, 
S.~Dhawan$^6$,   
H.~Do$^6$, J.~Egger$^5$, S.~Eilerts$^2$\cite{SE},    
H.~Fischer$^2$\cite{HF}, W.~Herold$^5$, 
V.~V.~Issakov$^4$, H.~Kaspar$^5$, D.~E.~Kraus$^3$, 
D.~M.~Lazarus$^1$, 
P.~Lichard$^3$, J.~Lowe$^2$, J.~Lozano$^6$\cite{JL},   
W.~Majid$^6$\cite{WMa}, W.~Menzel$^7$\cite{WMe},  
S.~Pislak$^{8,6}$, A.~A.~Poblaguev$^4$, P.~Rehak$^1$,  
A.~Sher$^3$ J.~A.~Thompson$^3$, 
P.~Tru\"ol$^{8,6}$, and M.~E.~Zeller$^6$   
}

\address{
$^1$ Brookhaven National Laboratory, Upton, NY 11973, USA\\ 
$^2$Department of Physics and Astronomy, 
University of New Mexico, Albuquerque, NM 87131, USA\\
$^3$ Department of Physics and Astronomy, University of Pittsburgh,
Pittsburgh, PA 15260, USA \\ 
$^4$Institute for Nuclear Research of Russian Academy of Sciences, 
Moscow 117 312, Russia \\
$^5$Paul Scherrer Institut, CH-5232 Villigen, Switzerland\\ 
$^6$ Physics Department, Yale University, New Haven, CT 06511, USA\\
$^7$Physikalisches Institut, Universit\"at Basel, CH-4046 Basel, Switzerland\\
$^8$ Physik-Institut, Universit\"at Z\"urich, CH-8057 Z\"urich, Switzerland}
\date{October 21, 1999}
\maketitle

\begin{abstract}
More than 400 \kpimumu\  events were observed in a rare $K^+$ decay
experiment at the AGS.  
Normalized to the \ktau\ decay, the branching ratio is 
determined to be  $(9.22 \pm 0.60 (stat) \pm 0.49 (syst) )\times 10^{-8}$. 
This branching ratio and the $\mu\mu$ mass spectrum is in very good
agreement with the measurement of the \kpee\ decay, but deviates
significantly from the previous measurement. 

\end{abstract}

}
In the Standard Model, the decay $K^+\rightarrow \pi^+\mu^+\mu^-$
($K_{\pi\mu\mu}$) proceeds through the same mechanism as
$K^+\rightarrow \pi^+e^+e^-$ ($K_{\pi ee}$). It has been recognized
for a long time that both decays are dominated by long distance
contributions
involving one photon exchange 
\cite{vainshtein,eilam,bergstroem,lichard}
and therefore can be
described by a vector interaction form factor. 
Consequently, 
the ratio of the two
decay rates depends only on the shape of this form factor.
Comparisons of the \pmm\, and \pee\, measurements 
are of great interest, not only within
the framework of the Standard Model, but also because 
differences in the form factors measured separately in the two decays
would indicate new physics,
in particular those extensions that involve mass dependent couplings. 
The dilepton mass range below 350 MeV/$c^2$ is almost inaccessible
by other means. 
A precise measurement of the decay amplitude is also essential for 
future investigations of \pmm\, in the context of P- and CP-violation
through measurements of the $\mu$ polarization, as suggested by many 
theoretical calculations \cite{geng}. 

The $K_{\pi ee}$ mode has been measured several times since the
mid-70's 
\cite{bloch,alliegro,deshpande},
most recently with high precision by this group
\cite{e865piee}, which observed a sample of 10,000 
events. 
This recent measurement firmly established 
that the decay proceeds through  a vector interaction 
and provided 
a precise measurement of the form factor. 
The result can be quoted as, 
$Br(\pi e e)=[2.94\pm 0.05\,(stat.)\pm 0.13\,(syst.)
\pm 0.05\,(model)]\times 10^{-7}$
and $\delta  =  2.14\pm 0.13\,(stat.)\pm 0.15\,(syst.)$ where $\delta$
is a parameter in the form factor, 
$f_V(z)=f_0(1+\delta z);\;z\equiv M_{\ell\ell}/m_K$ with $\ell=e$ or $\mu$
\cite{fv}.
The slope of the form factor, which is significantly larger than the
prediction of 
meson
dominance models or leading order chiral perturbation theory, 
has a direct impact on the prediction of the \pmm\ branching ratio. 

The first observation of the $K_{\pi\mu\mu}$ decay
was reported by
the E787 collaboration at the AGS in 1997 \cite{adler} .
The data sample consists of 13 fully reconstructed
three track and 221 partially reconstructed
two track events, with an estimated background
of 2.4 and 25 events, respectively. 
The $\mu\mu$ mass spectrum was not analyzed.  The branching ratio
was measured to be 
$[5.0 \pm 0.4(stat) \pm 0.7(syst) \pm 0.6(th) ]\times 10^{-8}$.
The central value used the form factor shape measured by
\cite{alliegro}, corresponding to $\delta=1.31$, 
and the last error came from the theoretical uncertainty
of the form factor shape.  
This branching ratio measurement was found to be 2.2 $\sigma$ below the expectation from
the then existing \pee\ measurement \cite{alliegro}. 
The significance of the 
discrepancy becomes greater when this branching ratio is compared to
the new \pee\ measurement, which predicts a model independent 
\pmm\ branching ratio of
$(8.7\pm0.4)\times10^{-8}$.  
In view of the
important consequences of such a discrepancy, an experiment
capable of collecting a large \pmm\, sample with
fully reconstructed kinematics and 
improved systematic errors is called for.

The experiment reported here was performed at the Brookhaven Alternating
Gradient Synchrotron (AGS)  in 1997, using the E865 detector, which 
was designed to search for the lepton number violating
decay \kpimue\cite{bergman,pislak}.  
The data were collected in a period of six weeks.
The details of the detector design and performance 
are discussed elsewhere \cite{e865piee,e865nim}.  The components 
relevant to this
study are emphasized here. 

\begin{figure}[htb]
\epsfig{figure=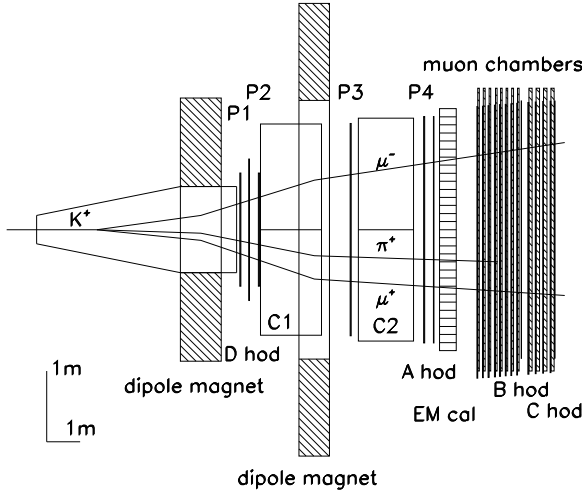,width=80mm}\centering
\caption[Plan view of the E865 detector]
{Plan view of the E865 detector. 
A \pmm\, decay is superimposed. 
}
\label{fig:detector}
\end{figure}

With 2.2$\times10^{12}$ protons on target, an
unseparated 6 GeV beam of $1.5\times10^{7}$ $K^+$, together with  
$3\times 10^{8}$ $\pi^+$ and protons, was produced per 1.6 sec AGS pulse. 
 Fig.~\ref{fig:detector} is a schematic diagram of the detector. Decay
products were measured in a proportional wire chamber based
 magnetic spectrometer system, together with
scintillating hodoscopes and an electromagnetic calorimeter.
Muons were identified in a 
range stack consisting of 24 planes of proportional 
tubes situated between iron absorber plates, and  two sets of hodoscopes
(48 pieces each) situated in the middle and at the end of the stack. 
The hodoscopes were segmented as vertical strips, separated in the
middle. 

The primary trigger is formed by the A and D hodoscopes
 and the electromagnetic  calorimeter,
where three charged particle hits are required.  Most of the trigger
rate, which is about 70,000/pulse,  comes from accidentals.  
This raw trigger rate is reduced by a prescale factor of
2000.    The most frequent $K^+$ decay that gives
three charged particles in the final state is the decay \ktau(\taus), 
and this decay mode is used for normalization. 
For \pmm\ decays, the next trigger level requires one muon on each side
of the detector.
Each muon is identified as a spatially
correlated  coincidence
between the B and C hodoscope hits.  The trigger rate at this stage is
about 2000/pulse, still dominated by accidental coincidences. 

\begin{figure}[htb]
\epsfig{figure=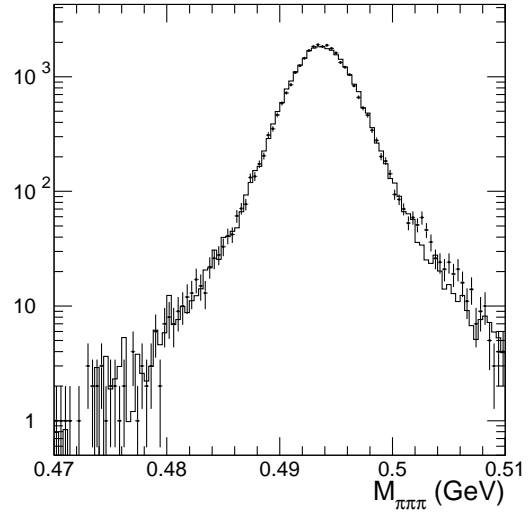,width=80mm}\centering
\caption[]
{Distribution of the reconstructed $\pi\pi\pi$ mass for \taus\
events. Points represent data, and the histogram is from a Monte Carlo
simulation. 
}
\label{fig:taums}
\end{figure}

In the off-line analysis, events are required to have three
reconstructed tracks that come from a common vertex in the decay volume, 
a reconstructed kaon momentum  consistent with the beam phase
space, and a timing spread between the tracks consistent with the
resolution.  Fig.~\ref{fig:taums} shows the reconstructed kaon mass
for \taus, in comparison with the Monte Carlo simulation. 
The mass resolution is $\sigma=2.2$ MeV.  The Monte Carlo
simulation, using the GEANT package,  
takes into account the detector geometry, particle decays and
interactions with the
detector, as well as the independently measured 
efficiencies 
for each detector component.  It reproduces all the essential
distributions of the normalization sample, \taus. 

\begin{figure}[thb]
\epsfig{figure=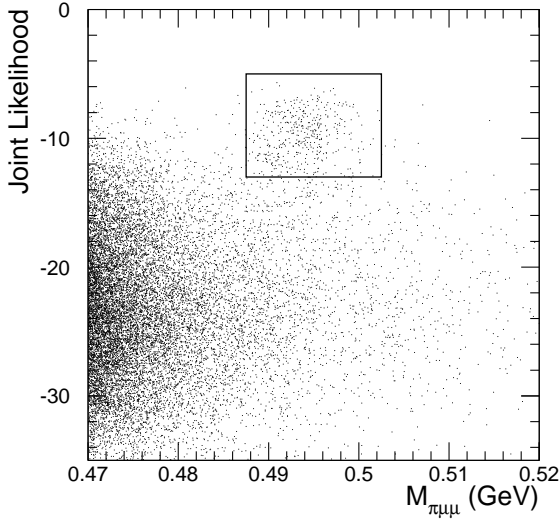,width=80mm}\centering
\caption[]
{Scatter plot of the joint likelihood vs 
$M_{\pi\mu\mu}$ for \pmm\ candidates. 
The box indicates the cuts for selecting \pmm\, events. 
}
\label{fig:pmmscat}
\end{figure}

\begin{figure}[htb]
\epsfig{figure=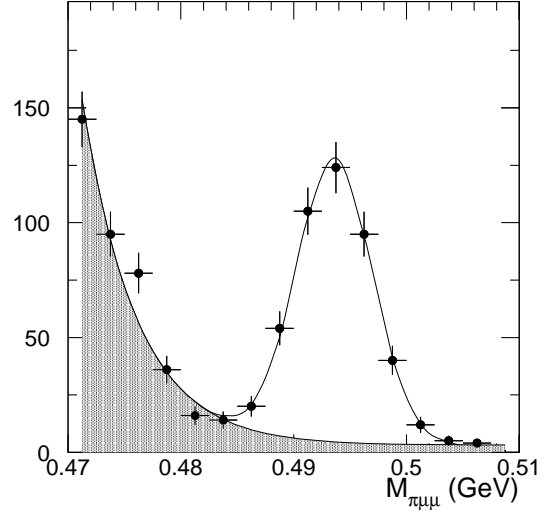,width=80mm}\centering
\caption[]
{The $\pi\mu\mu$ invariant mass distribution after requiring 
the joint likelihood to be greater than -13. The shaded area is the
background according to the fit.
 }
\label{fig:pmm_ms1}
\end{figure}

For a muon to reach the C hodoscope at the end of the muon stack, its
momentum must exceed 
1.3 GeV.  
Because of the trigger requirement, 
only events with 
the $\mu^-$ on the left side 
and the $\mu^+$ on the right side of the
detector are accepted. 
A muon is required to have deposited an energy in the calorimeter 
 consistent with a minimum ionizing particle, and to have 
B and C hodoscope hits  and sufficient  muon chamber hits along the 
projected trajectory through the muon stack.
At this stage, the accidental
background is virtually eliminated, and the $\pi\mu\mu$
candidates are
dominated by the background from  \taus, followed by
the $\pi^\pm\rightarrow\mu^\pm\nu$ decay.  The majority of such background
events have a $\pi\mu\mu$ mass much smaller than $M_K$.
To significantly suppress this background while keeping a high efficiency
for the signal, a joint likelihood function is constructed, using the
vertex quality, $K^+$ phase space, and the $\chi^2$ of the tracks.
Fig.~\ref{fig:pmmscat} is a scatter plot of the joint likelihood vs 
the $\pi\mu\mu$ invariant mass.  The cluster of events with high
likelihood value at the kaon mass are the \pmm\ signal, while the events
below the kaon mass with low likelihood value are background from 
\taus.   Monte Carlo simulations show that it is the pion decays in the
spectrometer magnet that cause significant mis-measurements of the 
particle momenta and spread the $\pi\mu\mu$ mass upwards to the signal
region.  The cuts based on the likelihood  
reduce such background. 
Fig.~\ref{fig:pmm_ms1} shows the $\pi\mu\mu$ invariant mass distribution
after requiring the joint likelihood function to be greater than -13. 
The observed
background events below the kaon peak are consistent with the simulation
of the background from \taus.  

	To subtract the background, we model the \pmm\ mass spectrum with 
\begin{eqnarray}
 n(x) &=& e^{p_1+p_2x+p_3x^2}+ p_4 + p_5 e^{-(x-x_0)^2/(2 p_6^2)} 
\end{eqnarray}
where $p_1,p_2,p_3,p_4,p_5,p_6$ are parameters to be determined,
$x=M_{\mu\mu}$, and $x_0$ is the kaon mass.
The first two terms represent the background, which is subtracted in the
signal region.

The same likelihood function is calculated for
the normalization data sample (\taus), as well as Monte Carlo events of
\taus\, and
\pmm\ decays. To verify that the background subtraction procedure is
reliable, the cut on the likelihood function is varied from -18 to -10.
Although the number of signal events drops from 535 events to 230 events, 
and the background fraction improves from 38\% to 1\%, 
the branching ratio measurement remains constant within 7\%.
The cut of -13 is chosen such that the statistical error, including the
uncertainty in background subtraction, is a minimum.
There are 430 events in the signal region, with 28 background events. 
The $\pi\mu\mu$ mass resolution is $\sigma=3.3$ MeV. 

\begin{figure}[htb]
\centerline {
\epsfig{figure=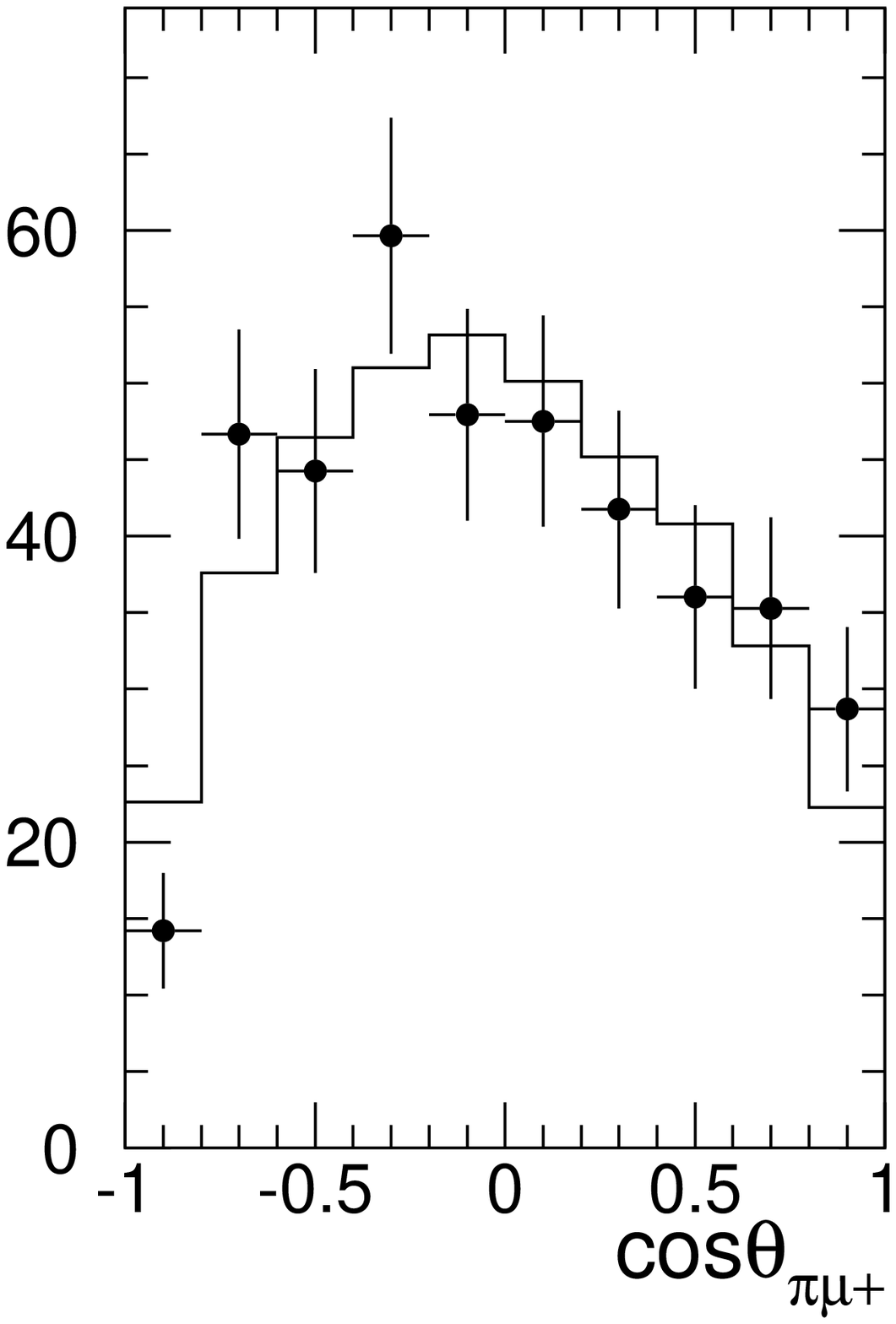,width=45mm,height=60mm} 
\epsfig{figure=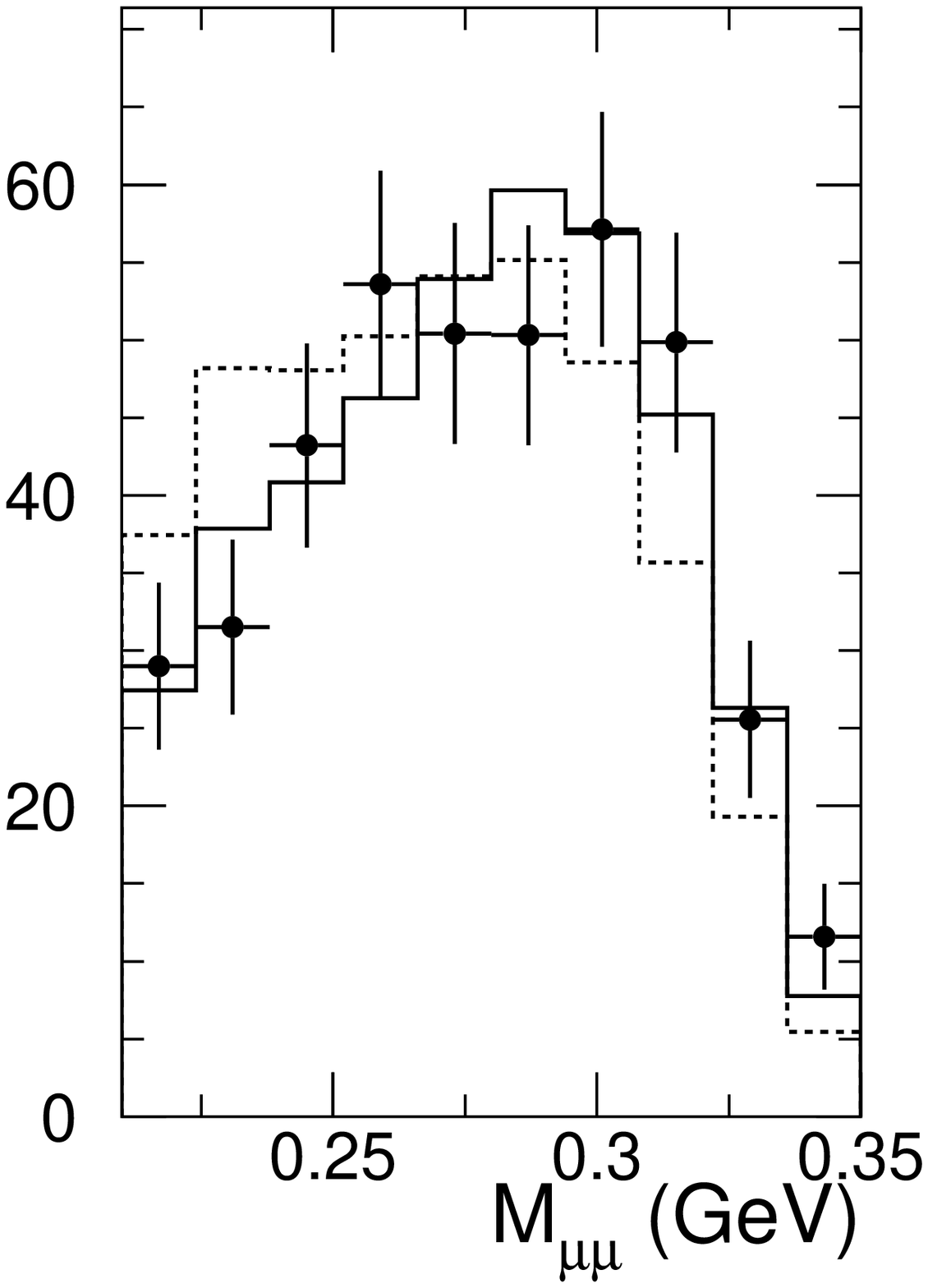,width=45mm,height=60mm}
}\centering
\caption[]
{Angular (left) and  $\mu\mu$ invariant mass (right) distributions
after background subtraction.
The points are data, the solid histogram is the result of a Monte Carlo 
calculation with $\delta=2.14$,
and the dashed histogram with constant form factor. 
}
\label{fig:mmms}
\end{figure}

To examine the decay mechanism, the $\cos\theta$ distribution 
(where $\theta$ is the angle between $\pi^+$ and $\mu^+$ in the center of
mass of the $\mu^+\mu^-$ pair), and $\mu^+\mu^-$ invariant mass
distribution are compared to the expectation, as shown in Fig
\ref{fig:mmms}.  The angular distribution is consistent with the decay
proceeding through a vector interaction, and the form factor
parameters describing the $M_{\mu\mu}$ spectrum are consistent with those
of \pee.

Normalized to the \taus\ branching ratio, the \pmm\ branching ratio 
and the form factor slope parameter are determined to be 
\begin{eqnarray}
 BR(\pi\mu\mu)& =& (9.22 \pm 0.60 )\times 10^{-8} \\ 
 \delta &=& 2.45 ^{+1.30}_{-0.95} 
\end{eqnarray} 
Errors are  statistical only, taking into account the uncertainty
in the background fluctuation.  

The similarity between the \taus\ and \pmm\ final state makes the
normalization very reliable.  The systematic error comes mainly from 
the uncertainty in the muon identification, and the background
subtraction procedure.  

\begin{figure}[htb]
\epsfig{figure=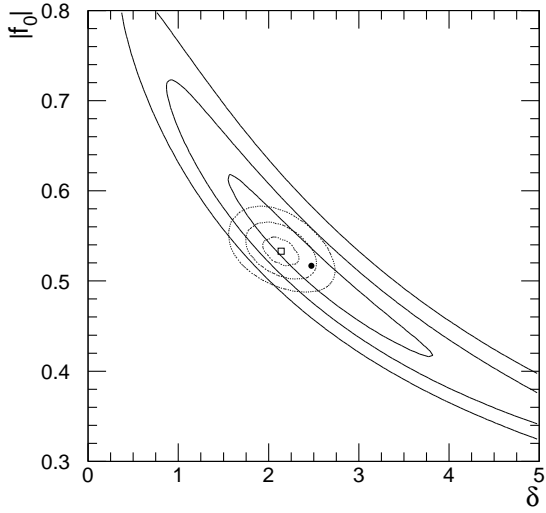,width=80mm}\centering
\caption[]
{ The $\chi^2$ contours of fits to the $M_{\ell\ell}$ distributions
of the \pmm\ data (solid),
and the \pee\ data(dashed).  The solid point is the $\chi^2$ minimum 
($\chi^2_{min}$) of the
$\pi\mu\mu$ fit, and the open point is that of the
$\pi ee$ fit.  The contours for each fit correspond to
$\chi^2=\chi^2_{min} + n$, $n=1,4$ or 9. 
Systematic errors are included.
}
\label{fig:f0lam}
\end{figure}
Table \ref{system} summarizes the systematic errors. 
There is sufficient overlap between the 
primary trigger and the $\pi\mu\mu$ trigger that the 
$\pi\mu\mu$ trigger
efficiency and primary trigger prescale factor can be 
precisely determined. 
To confirm the detector acceptance and efficiencies, 
many other checks were made. 
The Dalitz plot parameters of the \taus\, events
were found to be consistent with the published values.  
The muon identification efficiency 
was checked with 
the $K^+\rightarrow\pi^0\mu^+\nu$ decay followed by the \peeg\ decay.

The \pmm\ branching ratio divided by 
the model independent partial branching ratio of \pee\ with
$M_{ee}>0.15$ GeV is $0.458\pm0.043$. 
This is to be compared to the
expectation of 0.432 when $\delta=2.14$ is assumed. 
A more complete comparison 
is shown in Fig.~\ref{fig:f0lam}, where the $\chi^2$ contours 
are plotted as a function of $(f_0,\delta)$ for both decay modes.  
The consistency is evident. 

\begin{table}[htb]
 \begin{tabular}{l|c}
Sources      & $\sigma_{Br}/Br$   \\ 
\hline 
\taus\,branching ratio  &        0.01          \\
\taus\,prescale factor &   0.01            \\
$\pi\mu\mu$ trigger efficiency &   0.01            \\
B, C hodoscope efficiency & 0.02             \\
EM energy scale & 0.01            \\
$\mu$ Energy loss before C hodoscope & $ 0.03$ \\
Background subtraction \& cuts  & 0.03          \\
Reconstruction Efficiency  & $0.015$              \\
Magnetic field map     & $<0.005$         \\
\hline 
Total           & 0.053                      \\
\end{tabular}  
\caption[..]{ \it 
Systematic errors on the branching ratio measurement. 
}
\label{system}
\end{table} 

The two normalization samples, namely, \taus\, for \pmm\, and 
$K^+\rightarrow \pi^+\pi^0$ followed by \peeg\, for \pee, are 
compared and found to be consistent within the uncertainty. 

Our \pmm\, branching ratio measurement disagrees with 
that of \cite{adler}.  The discrepancy is 3.3 $\sigma.$ 
 
To summarize, we have measured the \pmm\ branching ratio to be 
$(9.22 \pm 0.60 (stat) \pm 0.49 (syst) )\times 10^{-8}$. Both  the decay
rate and the form factor shape are found to be in good agreement  with 
the expectation based on the \pee\ measurements,  
from which we conclude
that the mechanisms for these two decay modes are consistent with being
the same, as theoretically expected.

We gratefully acknowledge the contributions to the success of
this experiment by 
the staff and management of the AGS at the Brookhaven National
Laboratory, and the technical staffs of the participating institutions.
This work was supported in part by the U. S. Department of Energy, 
the National Science Foundations of the USA(REU program), 
Russia and Switzerland, and
the Research Corporation.


\begin{references}
\bibitem{fv} See \cite{e865piee} for the definition of the decay amplitude. 
Although a small, but significant, nonlinearity of the form factor was
found in the analysis of the \pee\, data\cite{e865piee}, 
the linear approximation is sufficient for this analysis.
\end{references}
\end{document}